\documentclass[12pt]{article}

\usepackage{amsmath}
\usepackage{amscd}
\usepackage{amssymb}
\usepackage{amsfonts}
\usepackage[mathscr]{euscript}
\usepackage{pstricks}
\usepackage{pst-node}
\usepackage[dvips]{pstcol}

\newcommand{\Tr}{\operatorname{Tr}}

\title{\bf Quantum Dynamical Semigroups and Non-decomposable Positive Maps}

\author{Fabio Benatti$^{a,b}$, Roberto Floreanini$^{b}$, 
Marco Piani$^{a,b}$\\
\small $^a$Dipartimento di Fisica Teorica, Universit\`a di Trieste,
Strada Costiera 11,\\
\small 34014 Trieste, Italy\\
\small ${}^b$Istituto Nazionale di Fisica Nucleare, Sezione di Trieste,
34100 Trieste, Italy} 

\date{\null}

\begin{document}

\maketitle

\vskip 2cm

\begin{abstract}
\noindent
We study dynamical semigroups of positive, but not completely
positive maps on finite-dimensional bipartite systems and
analyze properties of their generators in relation to non-decomposability 
and bound-entanglement. An example of non-decomposable semigroup leading to a
$4\times 4$-dimensional bound-entangled density matrix is explicitly obtained.
\end{abstract}

\section{Introduction}

Semigroups of dynamical maps are central
in the description of open quantum systems in weak interaction
with suitable external environments, acting as sources of dissipation and
noise. 
They have been succesfully used in many phenomenological applications in quantum
chemistry, quantum optics and statistical physics \cite{Spo,AL,BP,BF0}; 
they have also been applied
to dissipative phenomena induced by fundamental dynamics in various elementary
particle systems \cite{BF1,BFR1,BF2,BF3,BF4}.

In many instances, open systems relevant to physical applications can be modeled
as finite $d$-dimensional systems, whose states are described by 
density matrices $\rho$.
Their dynamics takes the form of semigroups of linear maps
$\gamma_t$, $t\geq 0$, satisfying the forward in time composition law 
$\gamma_t\circ\gamma_s=\gamma_{t+s}$, $t,s\geq 0$, and sending any initial
state $\rho$ into another state $\gamma_t[\rho]$ in the course of time; in
particular, $\gamma_t$ must be a positive map.

In line of principle, {\it positivity} is not sufficient to
guarantee full physical consistency of the maps $\gamma_t$: a more restrictive
property, namely {\it complete positivity}, needs to be imposed 
\cite{CH1,Tak,Kra}. 
This guarantees that not only the dynamics $\gamma_t$ of any system $S$ be 
positive, but that such is also the map $\gamma_t\otimes{\rm id}$,
with ``$\rm id$'' the identity operation, describing the time-evolution of $S$ 
statistically coupled to a generic inert $n$-level system $S_n$.

Complete positivity fully characterizes the form of the map $\gamma_t$
\cite{Tak,Kra} and, consequently, also its infinitesimal generator 
\cite{GKS,Lin}. 
On the contrary, if the map $\gamma_t$ is only positive,
then very little control is available either on its form or on that of its
generator: this partly justifies the reason why only
few examples of positive semigroups have been
considered so far in dissipative quantum dynamics 
\cite{AL,BP,Dum,Ell,BFBa}.

On the other hand, understanding the general structure of the set
of positive maps is becoming more and more important, 
since these maps serve as
entanglement witnesses in quantum information.%
\footnote{We refer to the review \cite{HHH0} and to the references therein
for more information concerning separability, entanglement and 
distillability.}
In this respect, 
particularly interesting from a physical point of view is to study the subset
of {\it decomposable} maps (see \cite{CH2,Sto,Wor,Kye1,Kos} and references
therein) which are sums
of a completely positive map and another completely positive map 
composed with the transposition; in fact,
the positive maps that are not decomposable are related
to the phenomenon of bound-entanglement and its 
non-distillability \cite{HHH0}.

Although both the mathematical and physical literature on non-decomposable
positive maps is rapidly growing, the question of decomposability of positive,
continuous semigroups, and not of generic maps, seems not to have so far
been raised.
The motivation for analysing this problem is twofold:
on one hand, the control of the mathematical structure of semigroups 
can be used to witness the presence of bound-entanglement, on the other, one 
may hope that they could shed some light on the process of bound-entanglement 
generation.

In the present investigation we will concentrate on how to reveal
bound-entanglement by means of quantum dynamical semigroups and
provide some general results relating positivity and 
decomposability of semigroups to relevant properties of their generators.
In the next section, we shall first briefly recall the notions of positivity,
complete positivity and decomposability when applied to one-parameter
semigroups.
Section 3 will be devoted to the study of positivity of product semigroups
on bipartite systems; some general results relating this condition to the
structure of the corresponding generators will be presented.
These considerations will then be used in Section 4 to analyze the structure 
of positive semigroups in relation to the notion of decomposability.
In particular, we shall examine in detail a bipartite system consisting of 
two $4$-dimensional systems whose dynamics is non-decomposable for an 
initial finite interval of time,
but becomes and stays decomposable afterwards.
We shall prove this by explicitly constructing a $4\times 4$-dimensional 
bound-entangled state that is witnessed by the initial non-decomposability 
of the dynamics.

\section{Complete positivity vs positivity}

According to the statistical interpretation of quantum mechanics, 
the physical states of a $d$-level quantum system $S$ are represented by density
matrices, that is by positive matrices 
$\rho\geq 0$, with $\Tr(\rho)=1$.
If we consider open sytems and describe their dissipative dynamics by 
semigroups of maps $\gamma_t$ sending any state $\rho$ at $t=0$ into 
$\gamma_t[\rho]$ at time $t>0$, then, for physical consistency, the
map $\gamma_t$ must preserve positivity, that is
$\gamma_t[\rho]\geq0$ at all times.

We shall focus on probability and positivity
preserving semigroups $\gamma_t$, continuous with respect to the trace-norm 
$\|X\|_1:=\Tr\sqrt{X^\dagger X}$ on the matrix algebra $M_d({\bf C})$.
From $\Tr(\rho_t)=1$ and
$\lim_{t\to0}\gamma_t={\rm id}$, it follows \cite{GKS} that 
$\gamma_t$ can be represented in exponential form, 
$\gamma_t={\rm e}^{t\,L}$, with the generator $L$ given by
\begin{equation}
\label{gen1}
L[\rho]=-i\Bigl[H\,,\,\rho\Bigr]
+\sum_{a,b=1}^{d^2-1}C^{ab}\Bigl(F_a\,\rho\, F^\dagger_b\,
-\,\frac{1}{2}\Bigl\{F^\dagger_bF_a\,,\,\rho\Bigr\}\Bigr)\ ,
\end{equation} 
where $H=H^\dagger\in M_d({\bf C})$, while the 
$F_a$, $a=1,2,\ldots, d^2-1$, are traceless $d\times d$ matrices 
forming together with $F_0:=\frac{{\bf 1}}{\sqrt{d}}$ an
orthonormal set in $M_d({\bf C})$:  
$\Tr(F_\mu^\dagger F_\nu)=\delta_{\mu\nu}$, $\mu,\nu=0,1,\ldots,d^2-1$.
In the following, it will prove convenient to isolate the so-called ``noise''
term
\begin{equation}
\label{noise}
N[\rho]=\sum_{a,b}\,C^{ab}F_a\,\rho\, F^\dagger_b\ ,
\end{equation}
from the pseudo-Hamiltonian contribution
\begin{equation}
\label{pseudo}
L_h[\rho]=-i\Bigl(H-\frac{i}{2}K\Bigr)\,\rho\ +\ 
i\rho\,\Bigl(H+\frac{i}{2}K\Bigr)\ ,\quad 
K=\sum_{a,b=1}^{d^2-1}\,C^{ab}\, F^\dagger_b F_a\ .
\end{equation}
\medskip

{\bf Remarks}

{\bf 1.1}\quad
A necessary and sufficient condition for the positivity of $\gamma_t$ is
espressed by the following constraint \cite{K}:
\begin{equation}
\label{Kcond}
\Tr\Big(P_i\,L[P_j]\Big)\, \geq\, 0\ ,\qquad i\neq j\ ,
\end{equation}
for all orthogonal resolutions $\{P_i\}$ of the identity: $\sum_iP_i={\bf 1}$,
$P_iP_j=\delta_{ij}P_i$.
 
{\bf 1.2}\quad
The $(d^2-1)\times(d^2-1)$ matrix $C$ of coefficients $C^{ab}$ is usually
called the Kossakowski matrix.
The condition (\ref{Kcond}) is too weak to fully characterize the matrix $C$; the
only general algebraic constraint following from (\ref{Kcond}) is
hermiticity: $C=C^\dagger$.
\medskip

As already observed, in analyzing open system dynamics, one usually asks 
for a more stringent condition than positivity, namely that the dynamical map
$\gamma_t$ be completely positive for all $t\geq 0$. 
Essentially, complete positivity guarantees that
the map $\gamma_t\otimes{\rm id}$ preserve the positivity of all states of
the compound system $S+S_n$, where $S_n$ is any $n$-level system.
As $S$ is assumed to be a $d$-level system, a theorem by Choi
\cite{CH1} ensures that the map $\gamma_t$ is completely positive iff 
the map $\gamma_t\otimes{\rm id}$ is positive for $n=d$.
The physical argument in support to the necessity of complete positivity
is that one cannot exclude that the system of interest $S$ might have
interacted with another $d$-level system in the past and become statistically
coupled to it.
In this case one should consider the two systems together,  
even though only one of them has a non-trivial time-evolution
$\gamma_t$, while the other one is dynamically inert \cite{Kra}.
The only states of $S+S_n$ that may develop negative eigenvalues under
$\gamma_t\otimes{\rm id}$, for $\gamma_t$ not completely positive, 
are the {\it entangled} ones, namely those which cannot be written in the 
separable form
\begin{equation}
\label{sepst}
\rho_{sep}=\sum_{ij}\lambda_{ij}\, \rho^1_i\otimes\rho^2_j\ ,
\end{equation}
where, $\lambda_{ij}>0$, $\sum_{ij}\lambda_{ij}=1$ and
$\rho^1_i$ and $\rho^2_j$ are any states of the two partner systems.
\medskip

One of the characteristic features of generic, trace-preserving
completely positive maps $\Lambda$
is that their structure is uniquely fixed: they can always be cast in the 
Kraus-Stinespring form \cite{Tak,Kra}
\begin{equation}
\label{cp1}
\Lambda[\rho]=\sum_\ell
V_\ell\,\rho\,V^\dagger_\ell\ ,
\end{equation}
where $V_\ell\in M_d({\bf C})$ and 
$\sum_\ell V_\ell^\dagger V_\ell={\bf 1}$.
Equivalently, they can be characterized by the following necessary and
sufficient condition \cite{HHH0} 
\begin{equation}
\label{cp2}
\Lambda\otimes{\rm id}[P^d_+]\geq 0\ ,
\end{equation}
where
\begin{equation}
\label{symmst}
P^d_+:=\frac{1}{d}\sum_{i,j=1}^d\vert i\rangle\langle j\vert
\otimes\vert i\rangle\langle j\vert\ ,
\end{equation}
is the symmetric projector with respect to a suitable
orthonormal basis $\{\vert j\rangle\}_{j=1}^{d}$ in ${\bf C}^d$.

Imposing (\ref{cp2}) on semigroups of maps $\gamma_t$, with generator as in
(\ref{gen1}), puts strong constraints on the corresponding
Kossakowski matrix $C$  (compare with Remark 1.2):
\medskip

{\bf Theorem 1} \cite{GKS,Lin}\quad
{\it
The map $\gamma_t$ generated by (\ref{gen1}) is completely positive iff  
$C$ is positive.}
\medskip

To decide whether a given mixed state $\rho$ of $S+S_n$ is entangled or not
is a rather subtle task; a very important tool is provided by the partial
transposition \cite{Per} $T\otimes{\rm id}$: indeed,
$T\otimes{\rm id}[\rho]$ can have negative eigenvalues only if $\rho$ is
entangled.
Having a non-positive partial transposition is sufficient for a
state to be entangled and is also necessary when $S$ is
a $2$-level system and $n=2,3$ \cite{HHH1}, but not in higher dimension where
there can be entangled states with positive partial
transposition: they are known as bound-entangled states 
\cite{H1,HHH0}. 
From a physical point of view, the entanglement contained in
bound-entangled states cannot be amplified by any local operation
on the two parties $S$ and $S_n$ \cite{HHH2,HHH0}.
An operational analytical approach to the description of
bound-entangled states is currently being elaborated which is based 
on the notion of {\it non-extendible orthonormal basis} \cite{Ter}.
Also, an attempt is being developed at formulating a thermodynamics of
entanglement (see the related references in \cite{HHH0}) where
bound-entanglement is the counterpart of heat. 

As already noticed, from a mathematical point of view, the fact that the 
algebraic structure of the Kossakowski matrix is fixed by complete positivity
and not by positivity is a consequence of the fact that, 
unlike completely positive maps, positive maps do not in general possess a 
structural characterization.  
The exception is provided by positive maps $\Lambda$ from $M_2({\bf C})$ to 
$M_{2,3}({\bf C})$, where the notion of {\it decomposability} \cite{CH2,Sto,Wor} 
fully characterizes them.
Indeed, any such $\Lambda$ can be written as
\begin{equation}
\label{decomp}
\Lambda=\Lambda_1+\Lambda_2\circ T\ ,
\end{equation} 
where $\Lambda_{1,2}$ are completely positive maps and $T$ is the 
transposition.
On the contrary, in higher dimension, there are positive maps that are not
decomposable \cite{CH2,Sto}, so that the decomposable ones form a
cone $\bf D$ strictly contained in the cone of all positive maps.
A useful tool in this context is offered
by the duality between ${\bf D}$ and the 
cone ${\bf T}$ of positive operators with positive partial
transposition \cite{Kye1}.
The duality is expressed
by the fact that $X\in{\bf T}$ if and only if  
\begin{equation}
\label{funct}
\langle\Lambda,X\rangle:=\Tr\Bigl(\Lambda\otimes{\rm id}[P^d_+]X^{T}\Bigr)
\geq 0
\end{equation}
for all $\Lambda\in{\bf D}$, where $X^{T}$ denotes transposition of $X$, and 
viceversa that $\Lambda\in{\bf D}$ if and only if (\ref{funct}) holds
for all $X\in{\bf T}$.
\medskip

{\bf Remark 2}\quad
Separable states belong to $\bf T$; further, if $\rho$ is separable, then 
every positive map $\Lambda$
is such that $\langle\Lambda,\rho\rangle\geq 0$.
Therefore, given a positive map $\Lambda$, if, for some $\rho\in{\bf T}$, 
$\langle\Lambda,\rho\rangle<0$, it follows that $\Lambda$ is non-decomposable 
and $\rho$ bound-entangled.

\section{Positivity of product semigroups}

In the following, we shall focus on one-parameter semigroups
$\{\Gamma_t\}_{t\geq 0}$ acting on states of the compound system $S+S$,
consisting of two independent, non-interacting copies of a $d$-dimensional
system $S$; the map $\Gamma_t$ can then be represented in product form,
$\Gamma_t:=\gamma^1_t\otimes\gamma_t^2$, with $\{\gamma^{1,2}_t\}_{t\geq 0}$ 
dynamical semigroups on $S$.
From the physical point of view, the semigroups
$\Gamma_t=\gamma^1_t\otimes\gamma_t^2$ have important applications since they
provide suitable dynamics describing the time-evolution
of two systems immersed in a same or in two different 
environments.

Let us first concentrate on the case $\gamma_t^1=\gamma_t^2$;
the result of Theorem 1 concerning 
the positivity of maps of the form $\Gamma_t=\gamma_t\otimes{\rm id}$ can then
be extended to semigroups of the form $\gamma_t\otimes\gamma_t$:
\medskip

{\bf Theorem 2} \cite{BFR3}\quad
{\it
The map $\Gamma_t=\gamma_t\otimes\gamma_t$ is positive iff the map 
$\gamma_t$ is completely positive, that is
iff the corresponding Kossakowski matrix $C$ is positive.}
\medskip

{\bf Remark 3}\quad
The fact that the map $\gamma_t$ must be completely positive for 
$\gamma_t\otimes\gamma_t$ to be positive provides more physical ground
to the necessity of complete positivity in open quantum dynamics with
respect to the usual argument based on $\gamma_t\otimes{\rm id}$.
Indeed, dynamical maps of the form $\gamma_t\otimes\gamma_t$
describe many relevant physical scenarios: among others,
decaying elementary particles in a noisy environment of gravitational origin
\cite{BF1,BFR1}, two entangled optical dipoles immersed in a heat bath
\cite{BF6,BFL}, or two entangled $2$-dimensional systems in random backgrounds
\cite{BFR4,BFR5}.
\medskip 

When the two parties do not exactly feel the same
environment, for instance because of local fluctuations of the heat
bath temperature, the appropriate description of their dissipative 
time-evolution is by means of maps $\Gamma_t=\gamma^1_t\otimes\gamma^2_t$, 
$\gamma^1_t\neq\gamma_t^2$.
The natural question is whether one can formulate general
necessary and sufficient conditions 
on $\gamma^{1,2}_t$ in order to guarantee the positivity of
the tensor product map $\Gamma_t$.
The question appears to be surprisingly complicated and no general answer is
yet available; some interesting results can nevertheless be formulated.
\medskip

{\bf Theorem 3}\quad
{\it
Let $\Gamma_t=\gamma^1_t\otimes\gamma^2_t$ and $C_{1,2}$ be the 
Kossakowski matrices corresponding to the semigroups $\gamma_t^{1,2}$ with
the generators as in (\ref{gen1}); for any invertible $V\in M_d({\bf C})$, 
let $\mathcal{V}$ be the invertible $(d^2-1)\times(d^2-1)$ matrix 
implementing the transformation
$V\,F^\dagger_a\,V^{-1}=\sum_{b=1}^{d^2-1}\mathcal{V}_{ab}F^\dagger_b$.
Then, the map $\Gamma_t=\gamma^1_t\otimes\gamma^2_t$ is positive only if 
$$
C:=C_1+\mathcal{V}^\dagger C_2\mathcal{V}\geq 0\ .
$$}
\smallskip

\noindent
{\bf Proof:}\quad 
The proof is a generalization of that of Theorem 2.
From Remark 1.1, a necessary and sufficient condition for the maps 
$\Gamma_t$ to be
positive is that, $\forall\, \vert\phi\rangle$ and 
$\vert\psi\rangle$ such that $\langle\phi\vert\psi\rangle=0$,
\begin{equation}
\label{proof1a}
\Big\langle\phi\Big\vert
L\Big[\vert\psi\rangle\langle\psi\vert\Big]\Big\vert\phi\Big\rangle\geq 0\ , 
\end{equation}
where $\displaystyle
L:=\frac{{\rm d}\Gamma_t}{{\rm d}t}\Bigl|_{t=0}=
L_1\otimes{\bf 1}_B+{\bf 1}_A\otimes L_2$, $L_{1,2}$ being the generators of
$\gamma_t^{1,2}$. 
Let $\Psi=[\psi_{ij}]$ and  $\Phi=[\phi_{ij}]$ be the $d\times d$
matrices given by the components of the orthogonal vectors 
$\vert\psi\rangle$, $\vert\phi\rangle\in{\bf C}^{d^2}$ with respect to a fixed
orthonormal basis $\{\vert i\rangle\otimes\vert j\rangle\}$, 
$i,j=1,2,\ldots,d$; 
then, $\Tr(\Phi\Psi^\dagger)=0$ and (\ref{proof1a}) reads
\begin{equation}
\label{proof1b}
0\leq\sum_{a,b=1}^{d^2-1}\Bigl(C_1^{ab}w^*_aw_b\,+\,C_2^{ab}v^*_av_b\Bigr)
\ ,\quad
w_a=\Tr\Bigl(F_a^\dagger\Phi\Psi^\dagger\Bigr)\ ,\quad
v_a=\Tr\Bigl(F_a^\dagger(\Psi^\dagger\Phi)^{T}\Bigr)\ .
\end{equation}
Let $W=\sum_{a=1}^{d^2-1} w_a F_a$ be a generic traceless
$(d^2-1)\times(d^2-1)$ matrix and $V$ a generic invertible matrix in 
$M_d({\bf
C})$, then choose $\Phi=VY^{-1}$ and $\Psi^\dagger=YV^{-1}W$
where $Y$ is the similarity matrix such that 
$YV^{-1}WVY^{-1}=\Bigl(V^{-1} W V\Bigr)^{T}$. The matrix $Y$ always exists 
since a given square matrix and its transpose have the same common 
divisors \cite{Gel}. It follows that $\Phi\Psi^\dagger=W$ and
$\Bigl(\Psi^\dagger\Phi\Bigr)^{T}=V^{-1} W V$, whence 
$0\leq \langle w\vert C_1+\mathcal{V}^\dagger
C_2\mathcal{V}\vert w\rangle$ for all vectors 
$\vert w\rangle\in{\bf C}^{d^2-1}$.
\hfill$\blacksquare$
\medskip

{\bf Remarks}

{\bf 4.1}\quad
When $C_1=C_2$, Theorem 3 reduces to Theorem 2, whereas, when 
$\gamma^1_t\neq\gamma_t^2$, it indicates that the map $\Gamma_t$ can
be positive without the maps $\gamma^{1,2}_t$ being both completely positive,
since $C_1$ and $C_2$ need not be both positive.
Theorem 3 gives a necessary, but hardly a sufficient condition: 
even positivity of the maps $\gamma^{1,2}_t$ may not be enforced.

{\bf 4.2}\quad
From a mathematical point of view, Theorem 3 can be used to construct
a rather rich class of positive maps that may be used to study the related
notions of non-decomposability and bound-entanglement: notice that this can be
achieved by controlling the generators $L$ which need not be positive.
 
{\bf 4.3}\quad
From a physical point of view, Theorem 3 might appear to
weaken the argument in favour of the necessity of complete positivity in open
quantum dynamics: a closer analysis reveals that it is not so.
In fact, when two copies of the same system are immersed in the same
environment, their time-evolutions $\gamma_t^{1,2}$ can differ only slightly;
then, a perturbative argument leads again to conclude that 
$\gamma^{1,2}_t$ must both be completely positive \cite{BFPR}. 
\medskip

As already noted positivity is harder to achieve than complete positivity;
the only general results are in $d=2$ \cite{K}.
They become particularly
simple to express with the additional request that the
Kossakowski matrix, $C$, be real and symmetric when
$\displaystyle F_a=\frac{\sigma_a}{\sqrt{2}}$, $a=1,2,3$,
where the $\sigma_a$ are the Pauli matrices.
In turn, this is equivalent to the fact that
the von Neumann entropy $S(\rho)=-\Tr(\rho\log\rho)$ never decreases
\cite{BFBa}. 
In this case, $C$ can always be chosen 
diagonal, $\displaystyle C={\rm diag}(c_1,c_2,c_3)$; 
if not, it can be diagonalized by orthogonal matrices that
can be used to define new Pauli matrices.
Then the map $\gamma_t$ is positive iff \cite{BFL,BFR4} 
\begin{equation}
\label{pos1}
c_1+c_2\geq0\ ,\quad c_2+c_3\geq0\ ,\quad
c_1+c_3\geq0\ ,
\end{equation}
whereas, according to Theorem 1, it is completely positive iff
$c_1\geq0$, $c_2\geq0$ and $c_3\geq0$.

In higher dimension $d\geq 3$, there are no general necessary and sufficient
conditions on the eigenvalues of $C$ that guarantee the positivity of
the corresponding semigroup. 
For example, we know that either $C_1$ or
$C_2$ may have negative eigenvalues and nevertheless lead to
a positive, but not completely positive, map 
$\Gamma_t=\gamma^1_t\otimes\gamma^2_t$. 

In order to explicitly construct examples of positive product semigroups, one
needs to supplement the necessary conditions of Theorem 3 with appropriate
sufficient ones; to this end we have the following result.
\medskip

{\bf Theorem 4}\quad
{\it
Suppose that the non-Hamiltonian terms in the generators of 
$\gamma^{1,2}_t$ are as follows,
\begin{equation}
\label{gen3}
D_{1,2}[\rho]=\sum_{\ell=1}^{d^2-1}c_{1,2}^\ell\Bigl(G_\ell^{1,2}\,
\rho\,G_\ell^{1,2}\,-\,\frac{1}{2}\Bigl\{(G_\ell^{1,2})^2\,,\,\rho\Bigr\}
\Bigr)\ ,\quad c^\ell_{1,2}\in{\bf R}\ ,
\end{equation}
where $G^{1,2}_\ell\in M_d({\bf C})$, together with 
$G^{1,2}_0=\frac{{\bf 1}}{\sqrt{d}}$, constitute two 
orthonormal sets of hermitian
traceless matrices.
Suppose that $c_1^\ell>0$ for all $\ell=1,2,\ldots,d^2-1$, and that 
$c_2^k=-|c_2^k|<0$, for one index $k$, while $c_2^\ell>0$ when
$\ell\neq k$; then, the map
$\Gamma_t=\gamma^1_t \otimes\gamma^2_t$ is positive if
$c_1^\ell\geq |c_2^k|$, $\ell=1,2,\ldots,d^2-1$ and
$c_2^\ell\geq |c_2^k|$, $\ell\neq k$. 
}
\smallskip

\noindent
{\bf Proof:}\quad
In this case, the right hand side of (\ref{proof1a}) can be recast as
\begin{eqnarray*}
I(\psi,\phi)&:=&\sum_{\ell=1}^{d^2-1}(c_1^\ell-|c_2^k|)
\Bigl|\Tr G^1_\ell\Phi\Psi^\dagger\Bigr|^2
\,+\,\sum_{\ell\neq k=1}^{d^2-1}
c^\ell_2\Bigl|\Tr G^2_\ell(\Psi^\dagger\Phi)^{T}\Bigr|^2\\ 
&+&\, |c_2^k|\Bigl(\sum_{\ell=1}^{d^2-1}
\Bigl|\Tr G^1_\ell\Phi\Psi^\dagger\Bigr|^2\,-\,
\Bigl|\Tr G^2_k(\Psi^\dagger\Phi)^{T}\Bigr|^2\Bigr)\ .
\end{eqnarray*}
Since $G^{1,2}_\ell$, $\ell=1,2,\ldots,d^2-1$, as well as
$\Phi\Psi^\dagger$ and $(\Psi^\dagger\Phi)^{T}$ are traceless and the
$G^{1,2}_\ell$ form a basis, we can expand $\Phi\Psi^\dagger=\sum_{\ell=1}
^{d^2-1}\Tr(G^1_\ell\Phi\Psi^\dagger)\, G^1_\ell$ and similarly for 
$(\Psi^\dagger\Phi)^{T}$ in terms of $G^2_\ell$. It thus follows that
$$
\sum_{\ell=1}^{d^2-1}\Bigl(\Tr G^1_\ell\Phi\Psi^\dagger\Bigr)^2=
\Tr\Bigl(\Phi\Psi^\dagger\Bigr)^2=
\Tr\Bigl(\Bigl(\Psi^\dagger\Phi\Bigr)^T\Bigr)^2
=\sum_{\ell=1}^{d^2-1}\Bigl(\Tr
G^2_\ell(\Psi^\dagger\Phi)^{T}\Bigr)^2 \ . 
$$
Further, extracting the $k$-th contribution and using the triangle 
inequality one gets
$$
\Bigl|\Tr G^2_k(\Psi^\dagger\Phi)^{T}\Bigr|^2\leq
\sum_{\ell=1}^{d^2-1}\Bigl|\Tr G^1_\ell\Phi\Psi^\dagger\Bigr|^2
\,+\,
\sum_{\ell\neq k=1}^{d^2-1}\Bigl|\Tr
G^2_\ell(\Psi^\dagger\Phi)^{T}\Bigr|^2\ ,
$$
which in turn implies 
$$
I(\psi,\phi)\geq\sum_{\ell=1}^{d^2-1}(c_1^\ell-|c_2^k|)
\Bigl|\Tr G^1_\ell\Phi\Psi^\dagger\Bigr|^2+
\sum_{\ell=1,\ell\neq k}^{d^2-1}(c_2^\ell-|c_2^k|)
\Bigl|\Tr G^2_\ell\Big(\Psi^\dagger\Phi\Big)^T\Bigr|^2\geq 0\ .
$$
\smallskip

\hfill$\blacksquare$
\medskip

{\bf Remark 5.}\quad
The class of generators of the form (\ref{gen3}) is not too narrow;
indeed, any generator (\ref{gen1}) with $F_a^\dagger=F_a$ and real 
symmetric Kossakowski matrices can be cast as in (\ref{gen3}):
one diagonalizes $C_{1,2}$ by means of
orthogonal transformations that are then used to turn the $F_a$'s into
$G^{1,2}_\ell$.
\medskip

{\bf Theorem 5}
{\it
In dimension $d=2$, if the maps $\gamma_t^{1,2}$ are both positive with 
corresponding real symmetric Kossakowski matrices $C_{1,2}$, then
the conditions of Theorem 4 are also necessary.}
\smallskip
 
\noindent
{\bf Proof:}\quad
As already
observed, we can assume the real symmetric matrix $C_1$
to be diagonal, while the matrix $C_2$ being also real symmetric can be
diagonalized by an orthogonal rotation $\widetilde{\mathcal{V}}$.
The latter always corresponds to a unitary transformation
$\sigma_a\mapsto V\sigma_a V^\dagger=
\sum_{b=1}^3\widetilde{\mathcal{V}}_{ab}\sigma_b$ 
of the orthonormal basis of Pauli matrices.
We can now use Theorem 3, with a unitary 
$\mathcal{V}=\mathcal{U}\widetilde{\mathcal{V}}$, where 
$\sum_{b=1}^3\mathcal{U}_{ab}\sigma_b=U\sigma_a U^\dagger$, 
with $U=\displaystyle\frac{\sigma_1+\sigma_2}{2}$ exchanging $\sigma_1$
with $\sigma_2$ and multiplying $\sigma_3$ by $-1$.
In this way $C_2$ is first diagonalized by $\widetilde{\mathcal{V}}$,
$C_2={\rm diag}(c_1^{(2)},c_2^{(2)},c_3^{(2)})$,
then its eigenvalues $c_{1,2}^{(2)}$ are exchanged while $c_3^{(2)}$
is left unchanged. Varying the Pauli matrices in $U$, from Theorem 3, one
derives $c^{(1)}_i+c^{(2)}_j\geq 0$, for $i,j=1,2,3$.
Therefore, only one of the Kossakowski matrices, say $C_2$, may
have negative eigenvalues. Moreover, condition (\ref{pos1}) enforces the 
presence of just one negative eigenvalue, whose absolute value must be smaller
than the other two.
\smallskip

\hfill$\blacksquare$

\section{Decomposability of positive semigroups}

As already remarked in the Introduction, positive maps that are not completely
positive are of great relevance in the physics of quantum information.
Much attention has lately been devoted to the study of positive maps that are
not decomposable, {\it i.e.} that cannot be cast in the form (\ref{decomp}), 
since they signal the phenomenon of bound entanglement.
In the light of the results of the previous sections, positive semigroups of the
form $\Gamma_t=\gamma_t^1\otimes\gamma_t^2$ can provide new insights in the
study of bound-entanglement since its appearance may be
put in relation to some characteristic features of the corresponding generators.
We shall first start by discussing an
explicit example and then prove some general results. 

As an instance of a map $\Gamma_t=\gamma_t^1\otimes\gamma_t^2$ which
is positive, but not completely positive, let us set $d=2$ and in 
(\ref{gen1}) choose $H=0$ and 
\begin{equation}
\label{pos3a}
C_1=\begin{pmatrix}1&0&0\cr0&1&0\cr0&0&1
\end{pmatrix}\ ,\qquad
C_2=\begin{pmatrix}1&0&0\cr0&-1&0\cr0&0&1
\end{pmatrix}\ .
\end{equation}
The generated semigroups have the explicit expressions
\begin{equation}
\label{pos3b}
\gamma^1_t[\rho]=\alpha\rho+\frac{1-\alpha}{2}\sigma_0
\ ,\quad
\gamma_t^2[\rho]=\rho-(1-\alpha)\rho_2\sigma_2\ ,\qquad
\alpha=\exp{(-2t)}\ ,
\end{equation}
where the expansion 
$\rho=\frac{1}{2}\sigma_0+\rho_1\sigma_1+\rho_2\sigma_2+\rho_3\sigma_3$
has been used, with $\sigma_0$ the $2\times 2$ unit matrix;
$\gamma_t^1$ is clearly completely positive, while $\gamma_t^2$
turns out to be only positive.

It is convenient to define the linear map 
$\Tr_2:M_2({\bf C})\mapsto M_2({\bf C})$,
where $\Tr_2[X]:=\Tr(X)\sigma_0$; it is completely positive
since it can be written in Kraus-Stinespring form as
\begin{equation}
\label{tr1}
\Tr_2[X]=\frac{1}{2}\sum_{\mu=0}^3\sigma_\mu X\sigma_\mu\ .
\end{equation}
Also, with respect to the standard representation of the Pauli
matrices, the transposition $T$ does not affect $\sigma_{1,3}$, while changes
the sign of $\sigma_2$, so that it can be explicitly written as 
\begin{equation}
\label{t1}
T[X]=\frac{1}{2}\Bigl(\sum_{\mu=0,\mu\neq 2}^3\sigma_\mu
X\sigma_\mu\,-\,\sigma_2X\sigma_2\Bigr)\ .
\end{equation}
With the help of (\ref{pos3b}), (\ref{tr1}) and (\ref{t1}),
$\Gamma_t=\gamma_t^1\otimes\gamma^2_t$ can be written as
\begin{equation}
\label{pos4}
\Gamma_t=\Bigl(\alpha\,{\rm id}_2\,+\,\frac{1-\alpha}{2}\Tr_2\Bigr)
\,\otimes\,\Bigl(\frac{1+\alpha}{2}{\rm id}_2\,+\,
\frac{1-\alpha}{2}T_2\Bigr)\ ,
\end{equation}
where, for the sake of clarity, the identity operation ${\rm id}_2$ on 
$M_2({\bf C})$
has been explicitly inserted.
\medskip

{\bf Remark 6}\quad
In (\ref{pos4}), the Kraus-Stinespring form (\ref{cp1}) is apparent 
in the first factor, while the second factor
is decomposed as in (\ref{decomp}).
Since the map $\Gamma_t$ from $M_4({\bf C})$
into itself is positive, but not completely positive, 
the question whether it is decomposable or not makes sense.
\medskip

We proceed by rewriting $\Gamma_t$ as $\Gamma^1_t+\Gamma^2_t\circ T_4$ where
$T_4=T_2\otimes T_2$ is the transposition on $M_4({\bf C})$ and 
$\Gamma_t^{1,2}$ are the following two linear maps on the same algebra,
\begin{eqnarray}
\label{G1}
\Gamma^1_t&=&\frac{1+\alpha}{2}\Bigl(\alpha\,
{\rm id}_2+\frac{1-\alpha}{2}\Tr_2\Bigr)\otimes {\rm id}_2\\ 
\label{G2}
\Gamma_t^2&=&\frac{1-\alpha}{2}\Bigl(\alpha\,T_2+
\frac{1-\alpha}{2}\Tr_2\Bigr)\otimes {\rm id}_2\ ,
\end{eqnarray}
where use has been made of the two identities
$T_2\circ T_2={\rm id}_2$ and $\Tr_2\circ T_2=\Tr_2$.
It turns out that the map $\Gamma^1_t$ is completely positive on
$M_4({\bf C})$ for any $t\geq 0$ for it is a tensor product of 
a sum of completely positive maps on $M_2({\bf C})$ with the identity.

In order to check whether the map $\Gamma_t^2$ is also completely positive, 
we use the criterion (\ref{cp2}) with
\begin{equation}
\label{symm4} 
P^4_+=\frac{1}{4}\sum_{a,b;c,d=1}^2\Bigl(\vert a\rangle\langle c\vert\otimes
\vert b\rangle\langle d\vert\Bigr)\otimes
\Bigl(\vert a\rangle\langle c\vert\otimes
\vert b\rangle\langle d\vert\Bigr)\ ,
\end{equation}
where $\vert a\rangle$, $a=1,2$ is a fixed orthonormal basis in ${\bf C}^2$.
Then, one explicitly finds
\begin{equation}
\label{G2cp}
\Gamma^2_t\otimes {\rm id}_4[P^4_+]=\frac{1-\alpha^2}{8}P^2_+\otimes
\begin{pmatrix}
1&0&0&0\cr0&0&0&0\cr0&0&0&0\cr0&0&0&1
\end{pmatrix}
\,+\,
\frac{1-\alpha}{8}P^2_+\otimes
\begin{pmatrix}
0&0&0&0\cr0&1-\alpha&2\alpha&0\cr0&2\alpha&1-\alpha&0\cr0&0&0&0
\end{pmatrix}\ ,
\end{equation}
where 
$P^2_+=\frac{1}{2}\sum_{a,b=1}^2\vert a\rangle\langle b\vert\otimes\vert
a\rangle\langle b\vert$.%
\footnote{
Notice that, for sake of compactness, we have chosen  
to represent the right hand side of
(\ref{G2cp}) as a tensor product which does not respect the splitting in 
(\ref{symm4}).}
It thus follows that $\Gamma^2_t\otimes {\rm id}_4[P^4_+]$ is positive
for $0\leq\alpha\leq1/3$, that is for $t\geq t^*$, 
$\displaystyle t^*:=(\log 3)/2$, while it has
a negative eigenvalue for $0<t<t^*$.
Therefore, $\Gamma^2_t$ is completely positive 
and $\Gamma_t$ decomposable for $t\geq t^*$.
\medskip

{\bf Remark 7}\quad
Because of the non-uniqueness of the decomposition (\ref{decomp}), the
above result does not necessarily mean that $\Gamma_t$ is not decomposable for
$0<t<t^*$: it may only indicate that the
maps $\Gamma^{1,2}_t$ in (\ref{G1}), (\ref{G2}) do not provide the right
decomposition in that range of time and that different ones have
to be looked for.
\medskip

To prove that the positive map $\Gamma_t$ in (\ref{pos4}) is
indeed non-decomposable for $0<t<t^*$, it is convenient to introduce the 
pure states $Z_{\mu\nu}$, $\mu,\nu=0,1,2,3$, of the form \cite{GKS}
\begin{equation}
\label{entst}
Z_{\mu\nu}=\bigg[{\bf 1}_4\otimes\Bigl(\sigma_\mu\otimes\sigma_\nu\Bigr)
\bigg]\,P^4_+\,\bigg[{\bf 1}_4\otimes\Bigl(\sigma_\mu\otimes\sigma_\nu\Bigr)
\bigg]\ ,
\end{equation}
constructed using the tensor products of Pauli matrices plus the identity 
$\sigma_0$.

Recalling the definition of the duality in (\ref{funct}), it can be 
checked that 
\begin{equation}
\label{wmn1}
W_{\mu\nu}:=\langle\Gamma_t,Z_{\mu\nu}\rangle=\frac{1}{4}
\Bigl(\alpha\delta_{\mu0}+\frac{1-\alpha}{4}\Bigr)\,
\Bigl[2(1+\alpha)\delta_{\nu0}+(1-\alpha)(1-2\delta_{\nu2})\Bigr]\ ,
\end{equation}
and, in particular,
\begin{equation}
\label{wmn2}
W_{02}=\frac{(\alpha-1)(1+3\alpha)}{16}\ , \quad
W_{11}=W_{23}=W_{31}=\,-\,W_{32}=W_{33}=\frac{(1-\alpha)^2}{16}\ .
\end{equation}
\bigskip

Let us then consider the following combination
\begin{eqnarray}
\nonumber
&&
\rho_{be}=\frac{1}{6}\Bigl(Z_{02}+Z_{11}+Z_{23}+Z_{31}+Z_{32}+Z_{33}\Bigr)\\
&&\nonumber\\
\label{rbe1}
&&
\!\!\!\!\!\!\!\!\!\!\!\!\!\!\!\!\!\!\!\!\!
=\frac{1}{24}\left(\begin{array}{cccccccccccccccc}
1&\cdot &\cdot &\cdot &\cdot &-1&\cdot &\cdot &\cdot &\cdot &-1&\cdot
&\cdot &\cdot &\cdot &1\\
\cdot &3&\cdot &\cdot &-1&\cdot &\cdot &\cdot &\cdot &\cdot &\cdot
&-1&\cdot &\cdot &-1&\cdot \\
\cdot &\cdot &1&\cdot &\cdot &\cdot &\cdot &-1&-1&\cdot &\cdot &\cdot
&\cdot &1&\cdot &\cdot \\
\cdot &\cdot &\cdot &1&\cdot &\cdot &1&\cdot &\cdot &1&\cdot &\cdot
&1&\cdot &\cdot &\cdot \\
\cdot &-1&\cdot &\cdot &3&\cdot &\cdot &\cdot &\cdot &\cdot &\cdot
&-1&\cdot &\cdot &-1&\cdot \\
-1&\cdot &\cdot &\cdot &\cdot &1&\cdot &\cdot &\cdot &\cdot &1&\cdot
&\cdot &\cdot &\cdot &-1\\
\cdot &\cdot &\cdot &1&\cdot &\cdot &1&\cdot &\cdot &1&\cdot &\cdot
&1&\cdot &\cdot &\cdot \\
\cdot &\cdot &-1&\cdot &\cdot &\cdot &\cdot &1&1&\cdot &\cdot &\cdot
&\cdot &-1&\cdot &\cdot \\
\cdot &\cdot &-1&\cdot &\cdot &\cdot &\cdot &1&1&\cdot &\cdot &\cdot
&\cdot &-1&\cdot &\cdot \\
\cdot &\cdot &\cdot &1&\cdot &\cdot &1&\cdot &\cdot &1&\cdot &\cdot
&1&\cdot &\cdot &\cdot \\
-1&\cdot &\cdot &\cdot &\cdot &1&\cdot &\cdot &\cdot &\cdot &1&\cdot
&\cdot &\cdot &\cdot &-1\\
\cdot &-1&\cdot &\cdot &-1&\cdot &\cdot &\cdot &\cdot &\cdot &\cdot
&3&\cdot &\cdot &-1&\cdot \\
\cdot &\cdot &\cdot &1&\cdot &\cdot &1&\cdot &\cdot &1&\cdot &\cdot
&1&\cdot &\cdot &\cdot \\
\cdot &\cdot &1&\cdot &\cdot &\cdot &\cdot &-1&-1&\cdot &\cdot &\cdot
&\cdot &1&\cdot &\cdot \\
\cdot &-1&\cdot &\cdot &-1&\cdot &\cdot &\cdot &\cdot &\cdot &\cdot
&-1&\cdot &\cdot &3&\cdot \\
1&\cdot &\cdot &\cdot &\cdot &-1&\cdot &\cdot &\cdot &\cdot &-1&\cdot
&\cdot &\cdot &\cdot &1
\end{array}\right)\ ,\\
&&\nonumber
\end{eqnarray}
where only the non-zero entries have been explicitly written down.
The matrix
$\rho_{be}$ is positive and normalized, and therefore represents a density
matrix; one further checks that it has positive partial transposition:
$T_4\otimes{\bf 1}_4[\rho_{be}]\geq 0$.
In addition, using (\ref{wmn2}), one finds
\begin{equation}
\label{rbe2}
\langle\Gamma_t,\rho_{be}\rangle=\frac{(1-\alpha)(1-3\alpha)}{48}\ .
\end{equation}
As the latter term becomes negative for
$\alpha>1/3$, that is for $0<t<t^*$, then, as explained in Remark 2,
the state $\rho_{be}$ cannot be separable.
As a consequence, $\rho_{be}$ is an explicit example of a
bound-entangled state in $4\times 4$ dimension; moreover, in the same interval
of time, $\Gamma_t$ provides a one-parameter family of non-decomposable maps.
\medskip

{\bf Remarks}

{\bf 8.1}\quad
The idea behind the construction of the bound-entangled state above is as
follows.
By expanding the pairing in (\ref{funct}) for
small $t$ with $X$ such that $\Tr(P^4_+\,X^{T})=0$, one finds that
only the ``noise'' term (\ref{noise})
actually contributes. 
Thus, in order to make 
$\langle\Gamma_t,X\rangle<0$, one may restrict to those convex
combinations $X$ of the projectors $Z_{\mu\nu}$ in (\ref{entst})
that have positive partial transposition ($X\in{\bf T}$), are orthogonal 
to $P^4_+$ and such that $\langle N,X\rangle<0$.
Indeed, in the case of (\ref{rbe1}) and (\ref{pos4}), one can check that
$\rho_{be}P^4_+=0$, whereas the noise term reads
$$
N=\Big(\Tr_2-\frac{{\rm id}_2}{2}\Big)\otimes {\rm id}_2\, + 
\,{\rm id}_2\otimes\Big(T_2-\frac{{\rm id}_2}{2}\Big)\ ,
$$
whence $\langle N\,,\,Z_{02}\rangle=-1/8$, 
while all the other pairings vanish, so that
$\langle N,\rho_{be}\rangle=-1/48$.

{\bf 8.2}\quad
The positivity of $\Gamma_t$ forces the noise term to fulfill
$\Big\langle\phi\Big\vert N[\vert\psi\rangle\langle\psi\vert]
\Big\vert\phi\Big\rangle\geq 0$ whenever
$\langle\phi\vert\psi\rangle=0$; this condition puts a strong constraint on 
the map $N$, but does not forces it to be positive.

{\bf 8.3}\quad
The state (\ref{rbe1}) is just one example of a class of bound entangled 
states witnessed by the map in (\ref{pos4}).
\medskip

Remark 8.1 suggests  that if the noise term (\ref{noise}) is positive and
decomposable, then the generated map $\Gamma_t$ is also decomposable.
\medskip

{\bf Theorem 6}\quad
{\it
If the noise term (\ref{noise}) in (\ref{gen1})
is positive and decomposable as in (\ref{decomp}), then the generated 
semigroup consists of decomposable maps.}
\smallskip

\noindent
{\bf Proof:}\quad
First, the composition $\Lambda\circ\Omega$ of two
decomposable positive maps $\Lambda=\Lambda_1+\Lambda_2\circ T$ and
$\Omega=\Omega_1+\Omega_2\circ T$, is decomposable. 
Indeed, since $T\circ T={\rm id}$, then
\begin{eqnarray*}
\Lambda\circ\Omega&=&\Lambda_1\circ\Omega_1+
\Lambda_2\circ \Bigl(T\circ\Omega_2\circ T\Bigr)\\
&+&
\Bigl(\Lambda_2\circ \Bigl(T\circ\Omega_1\circ T\Bigr)+
\Bigl(\Lambda_1\circ\Omega_2\Bigr)\Bigr)\circ T\ .
\end{eqnarray*}
Compositions and sums of completely positive maps are completely
positive; further, when $\Omega$ is a completely positive map, by
using (\ref{cp1}), one checks that also
the map $T\circ\Omega\circ T$ can be cast 
in the Kraus-Stinespring form, and 
is thus completely positive.
Therefore, the first line gives a completely positive map
and the second one a completely positive map composed with the transposition 
$T$.
It thus follows that $k$-times composition 
$N^k:=N\circ N\circ\cdots N$ of the noise in (\ref{noise})
are decomposable and thus also the exponential map 
$\displaystyle{\rm e}^{N\,t}=\sum_{k=0}^\infty\frac{N^k}{k!}$, $t\geq 0$.
The same is true of the strictly contractive completely positve map 
$\displaystyle{\rm e}^{L_h\,t}$,
$t\geq 0$, generated by the pseudo-Hamiltonian contribution (\ref{pseudo}).
Consequently, the
linear maps $\Bigl({\rm e}^{L_h\,t/n}\circ{\rm e}^{N\,t/n}\Bigr)^n$,
$n\geq0$, are decomposable too and, for all $X\in{\bf T}$, 
Trotter formula yields 
$$
\langle\Gamma_t,X\rangle=\lim_{n\to+\infty}\Tr\Bigl(
\Bigl({\rm e}^{L_h\,t/n}\circ{\rm e}^{N\, t/n}\Bigl)^n
\otimes\, {\rm id}[P_+]X^{T}\Bigr)\geq 0\ .
$$
\hfill$\blacksquare$
\medskip

{\bf Remarks}

{\bf 9.1}\quad
From Theorem 6, it follows that the noise term in the generator of $\Gamma_t$
in (\ref{pos4}) is not decomposable since the semigroup is not
decomposable for $0<t<t^*$.

{\bf 9.2}\quad
The condition of Theorem 6 is sufficient, but not necessary: a simple
counterexample to necessity is the semigroup of decomposable
positive maps $\gamma_t^2$ in (\ref{pos3b}).
It is easy to check that the noise term is not a positive map and therefore
not decomposable; for instance,
$\langle\phi\vert N[\vert\psi\rangle\langle\psi\vert]\vert\phi\rangle<0$
with $\psi=\phi=(1,i)$. 

{\bf 9.3}\quad
Interestingly, the non-decomposability of the positive map $\Gamma_t$ in
(\ref{pos4}) is a property that disappears after a finite
time-interval.%
\footnote{The same phenomenon
happens to entanglement: there are semigroups that transform initially 
entangled states into separable ones in a finite time \cite{D}} 
Similarly, a semigroup $\gamma_t$ may start positive and become completely
positive later.
Let $L$ be a generator of a semigroup on a
$2$-level system without Hamiltonian term and with Kossakowski matrix 
$$
C=\begin{pmatrix}
b&0&0\cr0&b&0\cr0&0&a-b\end{pmatrix}\ ,
$$ 
where $a,b>0$.
The positivity conditions (\ref{pos1}) are satisfied, whereas
$\gamma_t\otimes{\rm id}[P^4_+]$
has eigenvalues
$\displaystyle\mu(t)=1-{\rm e}^{-4bt}\geq 0$ and 
$\displaystyle\lambda_\pm(t)=1+{\rm e}^{-4b t}\pm\,2 {\rm e}^{-2at}$.
While $\lambda_+(t)$ is never negative, 
$\lambda_-(t)\geq0$ when $a\geq b$, whereas,
for $a<b$, $\lambda_-(t)$ is non-negative only for
$t\geq \hat{t}$, where $\hat{t}$ is such that
$\cosh 2b\hat{t}={\rm e}^{2
(b-a)\hat{t}}$; as a consequence, $\gamma_t$
is positive, but not completely positive for $0<t<\hat{t}$ and completely
positive for $t\geq \hat{t}$.

\section{Discussion}

The existence of bound-entanglement (and the related phenomenon of
non-distillability) 
asks for procedures able to recognize its presence.
In view of its relation to positive non-decomposable
maps, in the present investigation, we have focused on continuous 
one-parame-\break ter semigroups of positive maps,
which, unlike the ones consisting of completely positive maps, have so far 
been little considered in the literature.
The advantage of considering semigroups, instead of maps, is that 
their infinitesimal generators completely characterize their form,
which therefore can be controlled, at least to some extent.
In particular, we have analyzed bipartite finite dimensional systems and 
provided sufficient conditions 
on the structure of the generators giving raise to positive,
but not completely positive, product semigroups.

For such one-parameter families of maps, the question of whether they are
decomposable or not is of relevance both from the mathematical and physical
point of view; indeed, the property of decomposability is related, by duality,
to the class of states which possess a positive partial transpose, the
non-separable ones providing examples of the phenomenon of bound-entanglement.

In this context, we have constructed a product semigroup of positive,
non-decomposable maps witnessing a bound-entangled state
in $4\times4$ dimensions.
The construction is based on the analysis of the corresponding 
generator when acting on a particular class of states
naturally associated with it:
this strategy is different from the usual approach
relying on positive maps, since it is based on the study of an operator, 
the infinitesimal generator, that need not be (and indeed most of the times
is not) positive.

Finally, besides providing explicit non-decomposable maps and bound-entangled
states in $4\times 4$ dimensions, our construction also
indicates a more general procedure for generating similar examples in higher
dimensions.


\end{document}